# Spin-flop driven interfacial tunneling magnetoresistance in an antiferromagnetic tunnel junction

Xiaolin Ren[1], Ruizi Liu[*,2], Yiyang Zhang[1], Yuting Liu[2], Xuezhao Wu[2], Kun Qian[2], Kenji Watanabe[3], Takashi Taniguchi[4], Qiming Shao[*,1,2]

1 Department of Physics, The Hong Kong University of Science and Technology, Hong Kong SAR
2 Department of Electronic and Computer Engineering, The Hong Kong University of Science and Technology, Hong Kong SAR
3 Research Center for Electronic and Optical Materials, National Institute for Materials Science, 1-1 Namiki, Tsukuba 305-0044, Japan
4 Research Center for Materials Nanoarchitectonics, National Institute for Materials Science, 1-1 Namiki, Tsukuba 305-0044, Japan
* Corresponding emails: rliuap@connect.ust.hk; eeqshao@ust.hk

**The utilization of two-dimensional (2D) materials in magnetic tunnel junctions (MTJs) has shown excellent performance and rich physics. As for 2D antiferromagnets, the magnetic moments in different layers respond asynchronously and can be configured at various states under different magnetic fields, showing the possibility of efficient magnetic and electrical tunability. In this report, A-type antiferromagnetic (AFM) material $(Fe_{0.5}Co_{0.5})_5GeTe_2$ (FCGT) works as electrodes to realize full van der Waals magnetic tunnel junctions. Owing to the interfacial effect, the even-layer FCGT, although with zero net magnetization, exhibits spin selectivity in MTJ architecture contributing to a tunneling magnetoresistance (TMR) reaching about 25% at a low operating current 1 nA at 100 K and persists near room temperature. Due to the surface spin-flop (SSF) effect in antiferromagnetic FCGT, the alternation flexibility between the volatile and nonvolatile memory behavior is achieved. The interfacial TMR can be tuned efficiently in amplitude and even sign under different bias currents and temperatures. These findings show precise magnetoelectric manipulation in MTJs based on 2D antiferromagnets and highlight the promise of 2D antiferromagnets for spintronic devices.**

## 1. Introduction

Magnetic tunnel junctions, as the cornerstone of spintronic memory architecture, exploit tunneling magnetoresistance by typically controlling the relative alignment of magnetization states of ferromagnetic layers between parallel (low resistance) and antiparallel (high resistance) configurations. In recent years, the emergence of two-dimensional materials has heralded a new paradigm for MTJ due to their unique properties, such as atomically smooth interfaces and suppressed intermixing with neighboring materials[1,2]. Recently, breakthroughs, like giant TMR and good electrical tunability, have been made utilizing 2D magnets[3–5], however, it remains a challenge to achieve TMR at high temperatures due to their limited Curie temperatures[6–12].

Meanwhile, anti-ferromagnetic materials have attracted interest in spintronic studies owing to their excellent properties (e.g., ultrafast terahertz (THz) spin dynamics, good magnetic field immunity, and no identifiable stray fields)[13,14]. Recently, sizable TMR at room temperature has been realized based on 3D anti-ferromagnetic materials (e.g., $Mn_3Sn$[15,16], $Mn_3Pt$[17]), demonstrating the capability of AFM materials in MTJ construction. Furthermore, A-type van der Waals antiferromagnets have been extensively investigated. In these materials, interlayer antiferromagnetic coupling can stabilize an intermediate spin-flop state. In particular, this can manifest as a surface spin-flop, where low-field switching of the interfacial magnetic moment induces a considerable magnetoresistance[18–23]. Nonetheless, on the one hand, memory function is still confined to low temperatures (e.g., CrSBr[24–26], $CrI_3$[27,28]). On the other hand, owing to the bulk spin degeneracy[29–32], memory devices are either volatile, where the stored resistance states can be lost upon power and magnetic field removal. Or they rely on incorporation with another spin source[26,33–37].

In this work, we employ an A-type antiferromagnetic metal, $(Fe_{0.5}Co_{0.5})_5GeTe_2$, and construct the FCGT/$WSe_2$/FCGT van der Waals structure. Although FCGT exhibits bulk spin degeneracy, we observe that TMR depends on spin orientation in the layers interfacing $WSe_2$. Moreover, we find that the spin-flop effect plays a significant role in flipping the magnetization of surface layers in our antiferromagnetic MTJs. Meanwhile, both the amplitude and the sign of the TMR can be controlled by varying the bias current and temperature, demonstrating a high degree of tunability. Our results highlight the potential of constructing spintronic devices using 2D antiferromagnetic materials.

## 2. Result and discussion

Structurally, FCGT originates from ABC-stacked rhombohedral $Fe_5GeTe_2$ with the half substitution of the Fe atoms with Co and is reported to show perpendicular magnetic anisotropy (PMA)[38–41]. Its magnetic structure is defined by the intralayer ferromagnetic coupling and interlayer antiferromagnetic coupling, as demonstrated by the bilayer example in the inset of Figure 1a. To investigate the AFM behavior in FCGT, we first transfer odd-layer (~7 layers) and even-layer (~16 layers) flakes onto the Hall bar bottom electrodes (Figure 1a), respectively. Figure 1b and c show the anomalous Hall effect (AHE) data under out-of-plane (OOP) magnetic fields ($B$) from temperature $T = 10$ K to 350 K, displaying classical A-type antiferromagnetic performance with intralayer PMA[27,40–42]. The layer-parity-dependent magnetic behavior is reflected within the small $B$ region. For the odd-layer sample (Figure 1b), the presence of the square hysteresis loop, near zero field below $T = 300$ K, verifies its PMA nature and uncompensated net magnetization. On contrast, in the even-layer specimen (~16 layers, Figure 1c), akin to other A-type materials (i.e., $MnBi_2Te_4$[43,44]), Hall resistance ($R_{xy}$) shows a plateau with no hysteresis around zero field due to the compensated magnetization nature. When the magnetic field $B$ reaches the saturation field $H_s$ (i.e., > 5.5 T), $R_{xy}$ for both specimens saturate as all the magnetic moments align with the direction of $B$. In FCGT, both the intralayer ferromagnetic orders and the interlayer antiferromagnetic orders survive up to room temperature. Hysteresis is observed at large fields below 100 K, which may be due to the defects or vacancies[45]. Meanwhile, we introduce the one-dimensional chain model to understand the performance of layers under the magnetic evolution. Here, the optimized ratio $\frac{H_K}{H_J}$ for magnetic anisotropy $H_K$ and the interlayer exchange interaction $H_J$ is defined as 0.34[45] (Method). To model the magnetic transition behaviors of odd and even layers, we employ 3-layer and 4-layer FCGT as their respective representatives. The corresponding relations, regarding net magnetization ($M$) versus $B$, are displayed in Figure 1d and Figure 1e, respectively. It is emphasized that our calculations aim to capture the trend during magnetic evolution, where the deviations in the absolute values are expected due to the ideal consideration for simulation ($T = 0$ K, macrospin model and so on). The simulation results imply several distinct magnetic states in odd-layer and even-layer FCGT (Figure 3d-f, Supplementary Material 3). For better demonstration, we use the leftmost arrow as the representation of the interfacial magnetic moment. In small fields, odd-layer configurations exhibit a hysteretic net magnetic moment (↑↓↑ or ↓↑↓) (Figure 1d), whereas even-layer structures maintain a zero net moment (↑↓↑↓ or ↓↑↓↑) (Figure 1e). Under intermediate fields, surface spin-flop transitions (i.e., ↑↑↓↑) occur in even-layer FCGT. Proceeding through a canted AFM (cAFM) intermediary, when $B$ reaches $H_s$, all the moments in both odd and even layers transition to the FM configurations (i.e., ↑↑↑ and ↑↑↑↑). Note that, experimentally, 16-layer FCGT features no signature for SSF in Figure 1c, rather than the two-step spin-flop transition as simulated, which is attributed to the negligible influence on net magnetization in thick flakes[45].

We then establish FCGT1 (17L)/WSe$_2$ (4 L)/FCGT2 (50 L) MTJ, where the basic magnetic characterization of the flakes is shown in Figure S5, respectively. Figure 2a shows the sandwiched structure of our antiferromagnetic MTJ, whose optical image is displayed in Figure 2b. We determine the thickness of each 2D material layer through Atomic Force Microscope measurements. First, we conduct the electric characterization with zero magnetic field at 100 K. The nonlinear I-V behavior reveals the tunneling behavior at the FCGT/WSe$_2$ interface (Figure 2c). We next study the field dependence of MR under the out-of-plane magnetic field $B$ at a constant DC current bias of 1 nA at 100 K, as shown in Figure 2d. Beginning from an initial saturation at high negative field $B$ = -8 T (forward direction, blue curve), MR remains at a low resistance state ($R_L$) and nearly field-independent until an abrupt jump to a high stage ($R_H$) at about -0.5 T. Its $R_H$ state maintains up to around 2 T, above which the resistance drops back to $R_L$ state. The reversal of the $B$ sweeping direction (backward direction, red curve) grants a symmetric MR behavior, exhibiting $R_H$ state between $B$ = 0.5 T and $B$ = -2 T. According to Jullière model[46–50], the magnetoresistance ratio can be expressed $\text{TMR}\% = \frac{R_H - R_L}{R_L} \times 100\% = \frac{R_{AP} - R_P}{R_P} \times 100\%$, where $R_P$ ($R_{AP}$) refers to the parallel (antiparallel) state in MTJ model. The TMR ratio can be estimated to be around 25%, which is comparable to the value of current 2D ferromagnet-based MTJs[51–55]. Note that the abrupt change of TMR at -0.5 T (0.5 T) during forward (backward) magnetic field sweeping is striking, as the Hall measurements suggest, there is no abrupt change of net magnetization at ±0.5 T (Figure S5). It is a distinct feature unique from previous MTJs results[33–37,56,57].

Now, we discuss the physical origin of the "unexpected" TMR change at ±0.5 T. Classically, TMR effect comes from the imbalance in the spin-dependent density of states (DOS) near the Fermi level for electrodes[53,58–60]. Revised from Jullière model[46–50], the change in resistance $R$ can be understood by the distinct DOS change of tunneling electrons near electrode 1 ($DOS_1^{\uparrow/\downarrow}$) and electrode 2 ($DOS_2^{\uparrow/\downarrow}$) near the Fermi level. This can be expressed as $1/R \propto \sum_{\sigma=\uparrow,\downarrow} DOS_1^\sigma DOS_2^\sigma$. In terms of the A-type AFM with spin degeneracy[29–32], the major DOS change is expected during the modulation of the net magnetization (i.e., saturation of magnetization in AFM electrodes[56,57]), expected at around $B$ = ±5 T (Figure S5). However, the variation in TMR signal is subtle in this specific region (Figure 2d), indicating a non-dominated effect of bulk DOS change on the TMR response in our AFM MTJ. It distinguishes from the previous studies utilizing 2D antiferromagnets[33–37,56,57].

We then propose a mechanism that the TMR signals in antiferromagnetic FCGT-based MTJ mainly originate from the tunneling process through the $WSe_2$ barrier and adjacent FCGT monolayers (Figure 3). Recently, TMR has been observed in all-antiferromagnetic tunnel junctions (twisted bilayer CrSBr/bilayer CrSBr[26]), where different spin configurations at the interface result in different resistance states. Meanwhile, from theoretical calculation, the total spin transmission can be altered by aligning the relative direction of magnetic moments in two interfacial layers, although both antiferromagnetic electrodes have zero net magnetization, which can induce a sizable TMR in an all-antiferromagnetic junction[26,33,61]. Therefore, considering that the full transition from AFM state to FM state in AFM electrodes induces a negligible change in TMR response, the mechanism for TMR in our antiferromagnetic junction is most likely controlled by the relative magnetization of two interfacial FCGT layers[29,62]. Moreover, a sharp transition in the TMR is evident at the onset of the SSF state (Figure 2d), which is a direct signature of the spin orientation in the two interfacial layers from parallel to antiparallel, despite an insignificant change in $R_{xy}$ (Figure S4).

To further investigate the process of how TMR responds to the spin configuration of interfacial FCGT layers, we perform a series of TMR measurements by sequentially limiting the maximum applied magnetic field ($B_{max}$) Figure 3a-c. The simulation for odd- (3-layer) and even-layer (4-layer) FCGTs is incorporated to understand the corresponding spin dynamic pictures in FCGT1 and FCGT2, respectively (Figure 3d-f). All measurements are conducted after an initial saturation at $B$ = -8 T, and we focus on the resistance dynamics in the range (-2.5 T ~ 2.5 T) for better comparison (Figure 3a-c). In Figure 3a, under the forward sweeping direction (-8 T to 8 T, blue curve), the initialization forces all the magnetic moments align with B at first, and the parallel interfacial configuration contributes to $R_L$. Then $R_H$ state appears between around -0.5 T and 2 T, following the behavior as observed in Figure 2d, when $B$ sweeps along the forward direction. The experimental jump (~ -0.5 T) can be rationalized by the magnetic moment reorientation of the interfacial layer in FCGT2[63–66] (shown in the simulated spin dynamics for even-layer FCGT in Figure 3d), where the interfacial state for FCGT1 keeps ↓. Here, FCGT2 electrode starts the transition from SSF to AFM state (↓↓↑↓ to ↑↓↑↓ as seen in Figure 3e), which results in the antiparallel interfacial configuration. This antiparallel arrangement maintains until about 2 T, which coincides with the magnetization reversal in FCGT1 electrode (Figure 3f, Figure S5). Under backward sweeping with an initialization at $B$ = 8 T, similarly, $R_L$ switches to $R_H$ state at around 0.5T and drops back at about -2 T (red curve in Figure 3a). FCGT2 exhibits a different AFM configuration (↓↑↓↑) at $B$ = 0 T. As a result, during both forward and backward sweeping, TMR shows $R_H$ state at $B$ = 0 T. Distinct from the performance in Figure 3a, confining |$B_{max}$| to 2.5 T yields a nonvolatile TMR behavior. Although the forward sweeping process is identical to that

performed in Figure 3a, the magnetic states of FCGT2 retrace its path upon backward sweeping as shown in Figure 3e. Due to the absence of initialization at high positive magnetic field, FCGT2 maintains the same Néel vector of FCGT2 (↑↓↑↓) at $B = 0$ T during backward sweeping[63–66], thereby preserving the $R_L$ state without impeding the magnetization reversal in FCGT1 (here FCGT is ↑↓↑, Figure 3e-f). The increase in TMR occurs at -0.5 T, which stems from the reversible AFM/SSF conversion in FCGT2 (↑↓↑↓ to ↓↓↑↓). This reversible transition exclusively governs the resistance change when $|B_{max}|$ is further limited to 1.3 T, where MR curve shows no hysteresis (Figure 3c). We note that $B$ in the range ($|B| \leqslant 1.3$ T) is insufficient to induce magnetization changes in FCGT1 (interfacial magnetic moment is frozen at ↓)[63–66] (Figure 3f). Therefore, under a small $B$ range, TMR is volatile with only $R_H$ state at $B = 0$ T (Figure 3c). Combined with our analysis in Figure 3d-f, our measurements through different magnetic field protocols reveal that the TMR in our antiferromagnetic MTJ is strongly sensitive to the configuration of the interfacial layers.

To unravel the role of bias current in modulating interfacial spin polarization, the magnetoresistance was characterized across a range of bias currents (-1 μA to 1 μA), as shown in Figure 4a and 4b. Similar to the previously reported results based on 2D magnets, the generated interfacial spin polarization features reconfiguration on both sign and amplitude when subjected to bias currents[51–55]. The corresponding TMR ratios for our FCGT MTJ are summarized in Figure 4c. When the DC current bias is 1 nA, the TMR ratio at 100 K reaches a maximum of around 25%. At low bias currents ($|I| < 10$ nA), the parallel magnetic alignment in the layers near the barrier (FCGT1 and FCGT2) yields low resistance, and antiparallel alignment results in high resistance, which leads to a positive TMR. At higher bias, TMR gradually decreases and reverses the sign at ~ ±10 nA. As the amplitude of the current bias exceeds 50 nA, the difference between $R_P$ and $R_{AP}$ tends to be negligible. We also observe a similar trend in the results at 10 K (Supplementary Material 4). This is attributed to that the current amplitude modulates the disequilibrium in the density of states between spin-polarized tunneling electrons at the $WSe_2$ barrier interfaces[67–72] (Supplementary Material 5).

To investigate the influence of the temperature on tunneling behavior, we conduct temperature-dependent magnetoresistance measurements on FCGT-based magnetic tunnel junction (MTJ) devices. The MTJ maintains reliable tunneling operation over a temperature range from 10 K to 300 K (Supplementary Material 2). As shown in Figure 5a and 5b. The switch between low and high MR states requires higher fields as temperature decreases, whose TMR under various temperatures is extracted in Figure 5c. The spin valve effect can be maintained up to near room temperature (~290 K, Supplementary Material 6). A maximum TMR exceeding 30% is achieved at 10 K, followed by a monotonic decrease with rising temperature. Its physical mechanism can be attributed to the suppression in spin polarization caused by increased temperature[73]. In MTJs, the interfacial spin polarization from both electrodes, ($P_{i1}$ and $P_{i2}$, respectively) can be evaluated with $\text{TMR}\% = \frac{2P_{i1}P_{i2}}{1-P_{i1}P_{i2}} \times 100\% \approx \frac{2P_i^2}{1-P_i^2}$, where the electrodes with the same materials can give an approximated averaged evaluation by considering the same spin polarization ($P_{i1}P_{i2} \to P_i^2$) [52,53,74]. Figure 5d systematically analyzes the derived temperature-dependent interfacial polarization values. The spin polarization monotonically weakens from about 38% at 10K with increasing temperature. Its qualitative relation can be understood with Bloch's law[10,49,75]: $P_i(T) = P_0(1 - \alpha T^{3/2})$, where $\alpha$ is the temperature-dependent polarization decaying material constant and $P_0$ is spin polarization at 0 K. Upon fitting, our $\alpha$ is $1.16 \times 10^{-4}$ $K^{-2/3}$. It is comparable with reported in other 2D ferromagnetic materials, such as $Fe_3GeTe_2$[75] and $Fe_3GaTe_2$[51].

## 3. Conclusion

In summary, this study demonstrates the surface spin-flop transition driven in van der Waals antiferromagnetic FCGT MTJ. In the FCGT/WSe$_2$/FCGT heterostructure, even-layer FCGT exhibits robust tunable interfacial spin polarization in a wide temperature range. By confining the magnetic field range up to room temperature, the system allows versatile alternation between nonvolatile and volatile functionality. The sign and magnitude of the interfacial spin polarization are highly tunable via bias current. Our work highlights the application of interfacial spin polarization and surface spin-flop in antiferromagnetic MTJ architectures.

## 4. Methods

*Sample and device fabrication*: The process for fabricating FCGT/WSe$_2$/FCGT MTJ devices began with the acquisition of high-quality van der Waals (vdW) bulk single crystals of FCGT, hBN, and WSe$_2$ from HQ Graphene. We first exfoliated FCGT flakes onto on SiO$_2$ (285 nm)/Si substrates using adhesive tape. The Pt electrodes were fabricated using standard e-beam lithography and ultrahigh vacuum magnetron sputtering techniques. Next, FCGT and WSe$_2$ flakes, carefully chosen for their thickness and shape, were transferred onto these pre-patterned electrodes. The 2D heterostructure was accomplished using a dry transfer method[51]. To protect the FCGT layer from oxidation, the entire heterostructure was encapsulated with a thick layer of hBN. Finally, the device was annealed at 120 °C for 30 minutes to improve the contact between the different layers and to remove any air bubbles that might have been trapped during the transfer process. It is noted that all transfer steps were performed inside a nitrogen-filled glovebox, where the levels of oxygen and water were kept below 0.01 ppm, ensuring a clean interface and preventing contamination.

*Spin valve measurements*: Transport properties at low temperatures and as a function of temperature were characterized using a Cryogen Free Measurement System (CFMS). Electrical measurements, specifically I-V curves and magnetoresistance switching, were obtained using a Keithley 6221 current source and a Keithley 2182 nanovoltmeter.

*Characterization*: The doping level was determined by energy-dispersive X-ray spectroscopy (EDS). The morphology and elemental mapping were characterized using field emission scanning electron microscopy (SEM, JEOL-6700F). The thickness of the FCGT and $WSe_2$ flakes was characterized by atomic force microscopy (AFM, Microscope_Dimension ICON, Shimadzu; Dimension EDGE, Bruker).

*Simulations-Macrospin simulation with 1D chain model*: To gain deeper insight into the evolution of magnetic states in A-type antiferromagnetic FCGT during tunneling processes, we introduce a one-dimensional chain model. Here we suppose that intralayer ferromagnetic coupling exceeds interlayer antiferromagnetic coupling in strength[45]. The internal magnetic energy of an N-layer system with macroscopic spin can be expressed as:

$$U_N = \mu_0 M_s [\frac{H_J}{2} \sum_{i=1}^{N-1} \cos(\phi_i - \phi_{i+1}) - \frac{H_K}{2} \sum_{i=1}^{N} (\cos^2 \phi_i) - H \sum_{i=1}^{N} \cos\phi_i]$$

, where $\mu_0 M_s$ refers to the saturation magnetization per layer. The interlayer exchange interaction $H_J$ and magnetic anisotropy are quantified by the parameters $H_J = \frac{2J}{M_s}$ and $H_K = \frac{K}{M_s}$, respectively. $\phi_i$ denotes the angle of the macrospin in i-th layer with respect to the positive direction of the applied out-of-plane (OOP) magnetic field (z-axis). The optimized parameter for $\frac{H_K}{H_J}$ ratio is taken as 0.34[45]. After derivation, the equilibrium condition of i-th layer can be described as

$$\begin{cases} -\frac{H_J}{2}\sin(\phi_1 - \phi_2) + H_K\cos\phi_1\sin\phi_1 + H\sin\phi_1 = 0 & \text{(1st layer)} \\ -\frac{H_J}{2}\sin(\phi_i - \phi_{i-1}) - \frac{H_J}{2}\sin(\phi_i - \phi_{i+1}) + H_K\cos\phi_i\sin\phi_i + H\sin\phi_i = 0 & \text{(ith layer)} \end{cases}$$

*First-Principle simulation*: In the current study, the first-principles calculations have been performed by using the CASTE block. The Projector Augmented Wave (PAW) potential is applied to the elements, and the local density approximation (LDA) exchange-correlation function has been employed as it is more suitable than the generalized gradient approximation (GGA) for describing the magnetic properties of FCGT-like material. Then, the electrical and magnetic properties of FCGT was first investigated, the cut-off energy of the plane wave basis was set as 600 eV, and the convergence of the total energy was set to be less than $10^{-6}$ eV. All atomic positions and lattice constants of FCGT had been fully relaxed until the force on each atom was less than 0.01 eV Å$^{-1}$. The Brillouin zone sampling was performed by using the gamma-centered k-meshes 17 × 17 × 3, which is used for the self-consistent calculation and density of states calculation, respectively[56].

## Acknowledgement

The authors thank Prof. Dingfu SHAO for inspiring discussions. The authors at HKUST acknowledge funding support from NSFC/RGC Joint Research Scheme (No. N_HKUST620/21), National Key R&D Program of China (Grants No.2021YFA1401500), General Research Fund (Grant No. 16303521), and the State Key Laboratory of Advanced Displays and Optoelectronics Technologies. K.W. and T.T. acknowledge support from the JSPS KAKENHI (Grant Numbers 21H05233 and 23H02052) and World Premier International Research Center Initiative (WPI), MEXT, Japan.

## Conflicts of interest

The authors declare that they have no conflict of interest.

## Author Contributions

X. R. fabricated the devices, performed the measurement and analyzed the data with assistance from R. L. and Y. Z. Y.L. helped with the theoretical analysis with one-d macrospin model. X. W. helped to fabricate the platinum electrode. K. W., T. T. offer the high-quality hBN. Q. S. conceived and supervised the project. X. R. wrote the manuscript, and all authors discussed the content of the manuscript.

## Data Availability Statement

The data that support the findings of this study are available from the corresponding author upon reasonable request.

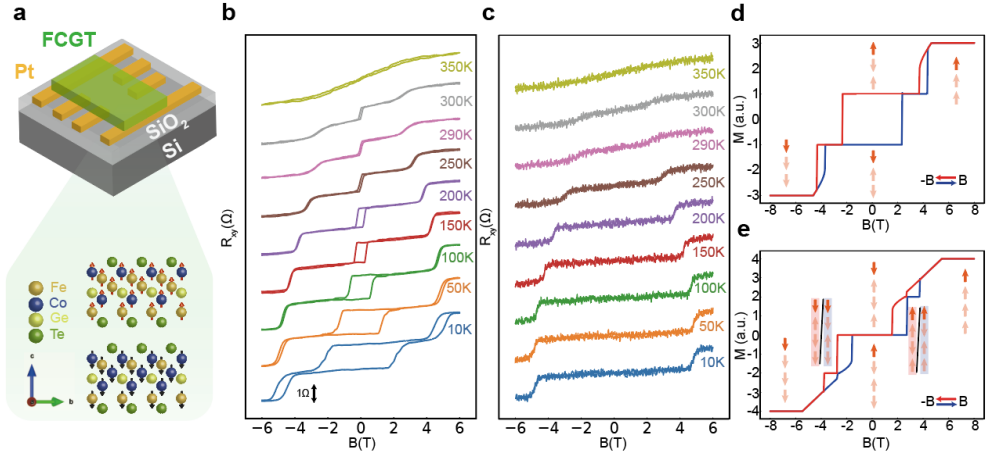

**Figure 1.** Basic magnetic properties of FCGT. (a)-Schematic diagram for FCGT device. The inset shows a side view of FCGT crystal structure. (b,c)-The Hall resistance ($R_{xy}$) along the c-axis under different magnetic fields for odd-layer (7 layers, b) and even-layer (16 layers, c) FCGT. (d,e)-Simulated magnetization for the odd-layer (3 layers, d) and even-layer (4 layers, e) FCGT with the corresponding spin textures at different out-of-plane magnetic fields based on a one-dimensional chain model. The net magnetization is normalized by the saturation magnetization of monolayer FCGT. The direction of each colored arrow corresponds to the alignment of magnetic moments in the system. The blue and red curves refer to the performance when the field sweeps along forward and backward directions, respectively. The corresponding SSF states are marked with blue and red background. Here the top layer for 3-/4-layer FCGT is considered to be the interfacial layer, which is highlighted in deep red.

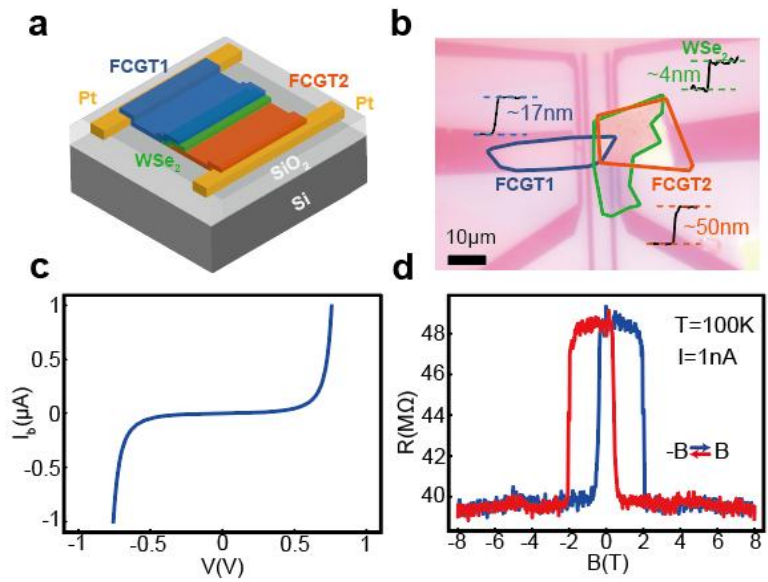

**Figure 2.** Characteristics of FCGT/WSe$_2$/FCGT MTJ device. (a)-Schematic diagram of the constructed antiferromagnetic MTJ. (b)-Optical image for the device and thickness for each flake. (c)-I-V characterization. (d)-TMR spectroscopy with current of 1 nA at 100 K when the magnetic field sweeps along forward (blue) and backward (red) direction.

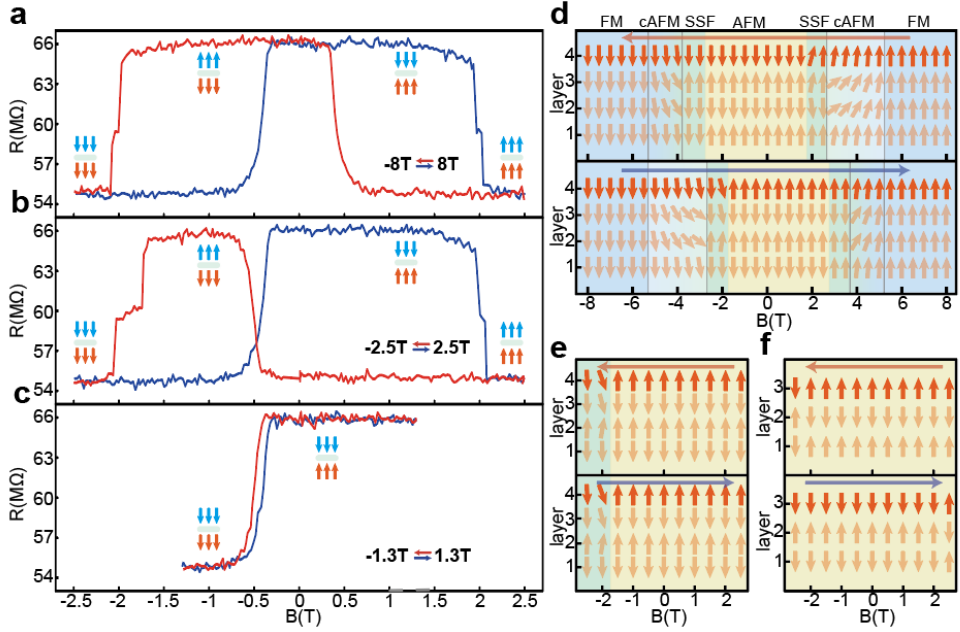

**Figure 3.** The measured TMR in sequence $-B_{max} \rightarrow B_{max}$ under the bias of 1 nA at 100 K. The magnetoresistance changes are recorded in blue (red) circles when the field sweeps along forward (backward) direction in the range of (a)- ±8 T, (b)- ±2.5 T, and (c)- ±1.3 T. The arrows in blue and orange refers to the interfacial magnetic situation in FCGT1 and FCGT2, respectively. Simulated magnetization configuration for odd- and even-layer flakes are summarized in (d-f). Here, the simulated results from a 4-layer FCGT are demonstrated in (d) and (e), for large field sweeping (i.e., $B_{max}$ = 8 T, d) and small field ($B_{max}$ = 2.5 T, e), respectively. The spin textures for 3-layer FCGT simulation with small field sweeping ($B_{max}$ = 2.5 T) are shown in (f). We highlight the top layer, considered as the interfacial layer. Colored arrows indicate the direction of magnetic moments within the system.

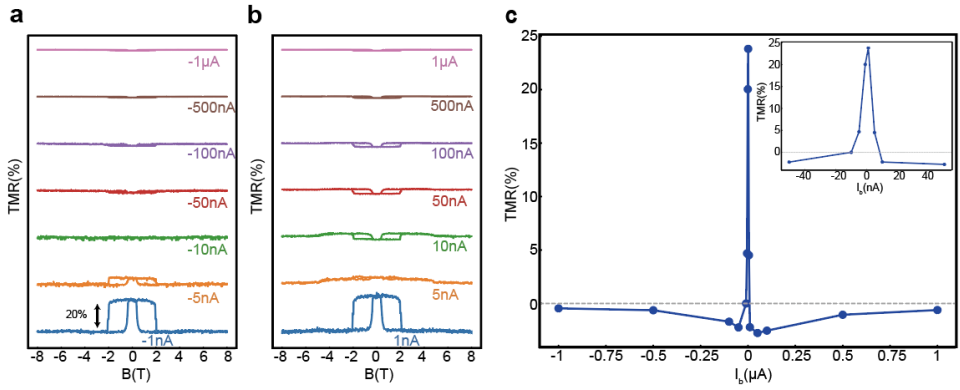

**Figure 4.** DC current dependence of TMR for FCGT-based MTJ (a),(b)-TMR measurement under different positive (a) and negative (b) bias currents at 100 K. (c)-bias-dependent TMR, which is extracted from (a) and (b) (inset is the rescaled diagram for c).

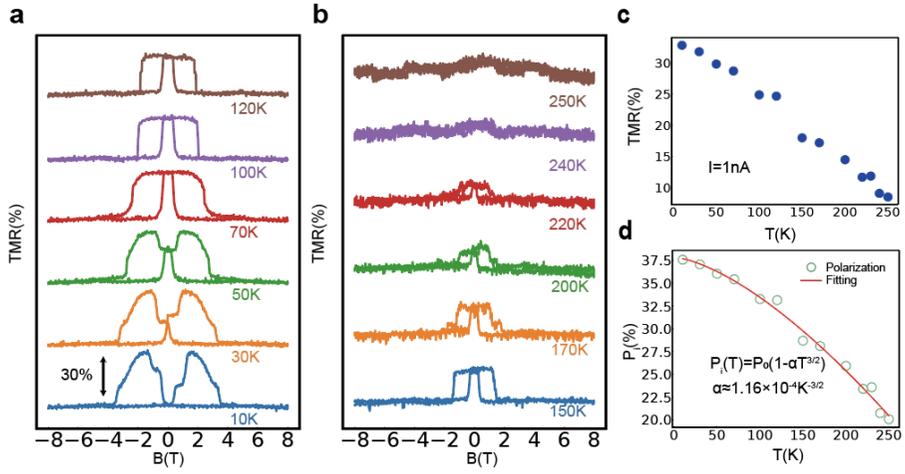

**Figure 5.** Temperature-dependent TMR in FCGT/WSe$_2$/FCGT MTJ. (a),(b)-TMR curves tested with 1 nA under temperatures varying from 10 K to 250 K. (c)-The extracted TMR versus temperature. (d)-The interfacial polarization ($P_i$) as a function of temperature (green hollow dots) is fitted (red line) into Bloch's law: $P_i(T) = P_0(1 - \alpha T^{3/2})$.